\documentclass[conference]{IEEEtran}

\usepackage{ifpdf}
\usepackage[nospace]{cite}
\usepackage{graphicx}

\usepackage{bm}
\usepackage[cmex10]{amsmath}
\usepackage{amssymb}
\usepackage{array}
\usepackage{syntonly}
\usepackage{amsthm}
\usepackage{amsfonts}
\usepackage{epsfig}

\begin{document}
\title{{\LARGE Decode-and-Forward Relay Beamforming with Secret and \\
Non-Secret Messages} }
%\vspace{3mm}
\author{
{\large Sanjay Vishwakarma and A. Chockalingam} \\
{\normalsize {Email: sanjay@ece.iisc.ernet.in, \ \ achockal@ece.iisc.ernet.in}} \\
{\normalsize Department of ECE, Indian Institute of Science, Bangalore 560012}
}

\maketitle

\begin{abstract}
In this paper, we study beamforming in decode-and-forward (DF) relaying 
using multiple relays, where the source node sends a secret message as well 
as a non-secret message to the destination node in the presence of multiple 
non-colluding eavesdroppers. The non-secret message is 
transmitted at a fixed rate $R_{0}$ 
and requires no protection from the eavesdroppers, whereas the secret message 
needs to be protected from the eavesdroppers. The source and relays operate 
under a total power constraint. We find the optimum source powers and weights 
of the relays for both secret and non-secret messages which maximize the worst 
case secrecy rate for the secret message as well as meet the information rate 
constraint $R_{0}$ for the non-secret message. We solve this problem for the 
cases when ($i$) perfect channel state information (CSI) of all links is known, 
and ($ii$) only the statistical CSI of the eavesdroppers links and perfect CSI 
of other links are known.
\end{abstract}
{\em keywords:}
{\em {\footnotesize
Cooperative relaying, physical layer security, secret and non-secret
messages, secrecy rate, multiple eavesdroppers. 
}} 
\IEEEpeerreviewmaketitle

\section{Introduction}
\label{sec1}
The foundation for secure communication using physical layer techniques was 
laid by Wyner in his work in \cite{ic0}, where the idea of secrecy rate and 
secrecy capacity for the wire-tap channel was introduced. The work was later
extended to the broadcast channel and the Gaussian channel in \cite{ic1} and 
\cite{ic2}, respectively. The broadcast nature of wireless transmissions makes 
them vulnerable to eavesdropping. Several works on secure wireless communication 
using single and multiple antennas have been reported in the literature, e.g., 
\cite{ic3,ic5,ic6, ic50}. Cooperative relays which act as distributed antennas can be 
used to improve the secrecy rate performance. Secrecy rates under cooperative 
relaying have been studied, e.g., \cite{ic7,ic8,ic9,ic11,ic12}. 

In \cite{ic1}, simultaneous transmission of a private message to receiver 1 
at rate $R_{1}$ and a common message to both the receivers at rate $R_{0}$ 
for two discrete memoryless channels with common input was considered. 
Recently, the work in \cite{ic1} has been extended to MIMO broadcast channel 
with confidential and common messages 
in \cite{ic20,ic21,ic22}. Motivated by the above works, in this paper, 
we consider communication of secret and non-secret messages between a 
source-destination pair, aided by multiple decode-and-forward (DF) relays, 
in the presence of multiple non-colluding eavesdroppers.
Both the secret and non-secret messages are intended for the destination. While 
the secret message needs to be protected from the eavesdroppers, the 
non-secret message need not be. The non-secret message is sent at a fixed 
rate $R_0$. There is a total power constraint on the source and relays powers. 
In this setting, our aim is to find the optimum source powers and relay weights 
(beamforming vectors) for the secret and non-secret messages. The objective is 
to maximize the worst case secrecy rate for the secret message and to meet the 
information rate constraint $R_{0}$ for the non-secret message. We solve this 
problem for two cases of channel state information (CSI) assumption. In the 
first case, perfect CSI of all links is assumed. In the second case, only the 
statistical CSI of the eavesdroppers links and perfect CSI of other links are 
assumed to be known.

$\bf{Notations:}$ $\boldsymbol{A} \in 
\mathbb{C}^{N_{1} \times N_{2}}$ implies that $\boldsymbol{A}$ is a complex 
matrix of dimension $N_{1} \times N_{2}$. $\boldsymbol{A} \succeq \boldsymbol{0}$ 
denotes that $\boldsymbol{A}$ is a positive semidefinite matrix. 
Transpose and complex conjugate transpose operations are denoted by 
$[.]^{T}$ and $[.]^{\ast}$, respectively. $\parallel.\parallel$ denotes 2-norm operation.
$\mathbb{E}[.]$ denotes the expectation operator.
%\vspace{-2mm}
\section{System Model}
\label{sec2}
\begin{figure}
\center
\includegraphics[totalheight=6.5cm,width=7.5cm]{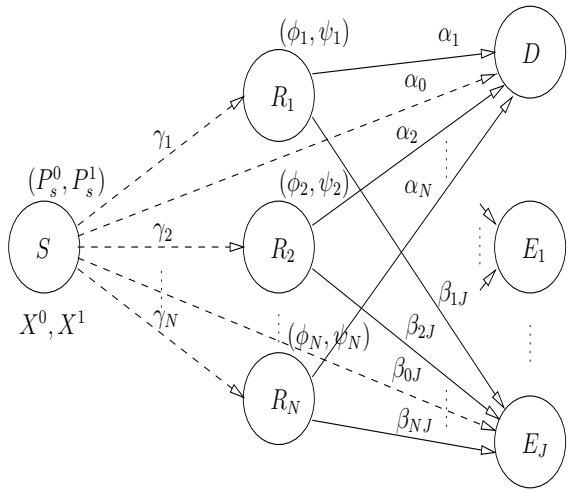}
\caption{DF relay beamforming with secret and non-secret messages.} 
\label{fig1}
%\vspace{-4mm}
\end{figure}

%\vspace{-0.0mm}
Consider a DF cooperative relaying scheme which consists 
of a source node $S$, $N$ relay nodes \{$R_1,R_2,\cdots,R_N$\}, an intended 
destination node $D$, and $J$ non-colluding eavesdropper nodes 
\{$E^{}_{1},E^{}_{2},\cdots,E^{}_{J_{}}$\}. 
The system model is shown in Fig. \ref{fig1}. In addition to the links from 
relays to destination node and relays to eavesdropper nodes, we assume direct 
links from source to destination node and source to eavesdropper nodes. The 
complex fading channel gains between the source to relays are denoted by 
$\boldsymbol{\gamma} = [\gamma^{}_1,\gamma^{}_2,\cdots,\gamma^{}_{N}] \in \mathbb{C}^{1 \times N_{}}$. Likewise, 
the channel gains between the relays to destination and the relays to $j$th 
eavesdropper are denoted by  
$\boldsymbol{\alpha} = [\alpha^{}_1,\alpha^{}_2,\cdots,\alpha^{}_{N}] \in \mathbb{C}^{1 \times N_{}}$ and 
$\boldsymbol{\beta}_{j} = [\beta^{}_{1j},\beta^{}_{2j},\cdots,\beta^{}_{Nj}] \in \mathbb{C}^{1 \times N_{}}$, 
respectively, where $j=1,2,\cdots,J_{}$. The channel gains on the direct 
links from the source to destination and the source to $j$th eavesdropper 
are denoted by $\alpha^{}_0$ and $\beta^{}_{0j}$, respectively. 

Let $P_T$ denote the total transmit power budget in the system (i.e., source 
power plus relays power). The communication between source $S$ 
and destination $D$ happens in two hops. Each hop is divided into $n$ channel 
uses. In the first hop of transmission, the source $S$ transmits two 
independent messages $W_{0}$ and $W_{1}$ which are equiprobable over 
$\{1,2,\cdots,2^{2nR_{0}} \}$ and $\{1,2,\cdots,2^{2nR_{s}} \}$, respectively. 
$W_{0}$ is the non-secret message to be conveyed to the destination at a fixed 
information rate $R_{0}$ which need not be protected from $E_{j}$s. $W_{1}$ is 
the secret message which has to be conveyed to the destination at some rate 
$R_{s}$ with perfect secrecy \cite{ic50}, i.e., $W_{1}$ needs to be 
protected from all $E_{j}$s. For each $W_{0}$ drawn equiprobably from the set 
$\{1,2,\cdots,2^{2nR_{0}} \}$, the source $S$ maps $W_0$ to an iid 
$(\sim \mathcal{CN}(0,1))$ codeword $X^{0}_{1},X^{0}_{2},\cdots,X^{0}_{n}$ 
of length $n$. Similarly, for each $W_{1}$ drawn equiprobably from the set 
$\{1,2,\cdots,2^{2nR_{s}} \}$, the source $S$, using a stochastic encoder,
maps $W_{1}$ to an iid $(\sim \mathcal{CN}(0,1))$ codeword 
$X^{1}_{1},X^{1}_{2},\cdots,X^{1}_{n}$ of length $n$. Let $P^{0}_{s}$ and 
$P^{1}_{s}$ denote the source transmit powers corresponding to the codewords
$X^{0}_{1},X^{0}_{2},\cdots,X^{0}_{n}$ and $X^{1}_{1},X^{1}_{2},\cdots,X^{1}_{n}$, 
respectively. In the $k$th $(1 \leq k \leq n)$ channel use, the source transmits
the sum of the weighted symbols, i.e., 
$\sqrt{P^{0}_{s}}X^{0}_{k} + \sqrt{P^{1}_{s}}X^{1}_{k}$. In the following, 
we will use $X^{0}$ and $X^{1}$ to denote the symbols in the codewords 
$X^{0}_{1},X^{0}_{2},\cdots,X^{0}_{n}$ and $X^{1}_{1},X^{1}_{2},\cdots,X^{1}_{n}$,
respectively.

In the second hop of transmission, relays retransmit the decoded
symbols $X^{0}$ and $X^{1}$ to the destination 
$D$. Let $\boldsymbol{\phi} = [\phi^{}_1,\phi^{}_2,\cdots,\phi^{}_{N}]^{T} \in \mathbb{C}^{N_{} \times 1}$ and 
$\boldsymbol{\psi} = [\psi^{}_1,\psi^{}_2,\cdots,\psi^{}_{N}]^{T} \in \mathbb{C}^{N_{} \times 1}$
denote the complex weights applied by the relays corresponding to the transmit symbols
$X^{0}$ and $X^{1}$, respectively.
The $i$th $(1 \leq i \leq N)$ relay transmits the 
sum of the weighted symbols which is $\phi_{i} X^{0} + \psi_{i} X^{1}$.

Let $y_{R_{i}}$, $y_{D_{1}}$, and $y_{E_{1j}}$ denote the received 
signals at the $i$th relay, destination $D$, and $j$th eavesdropper $E_{j}$, 
respectively, in the first hop of transmission. In the second hop of 
transmission, the received signals at the destination and $j$th eavesdropper
are denoted by $y^{}_{D_{2}}$ and $y^{}_{E_{2j}}$, respectively. We then have
\begin{eqnarray}
y^{}_{R_{i}} \ = \ \sqrt{P^{0}_{s}}{{\gamma}}^{}_{i}X^{0} + \sqrt{P^{1}_{s}}{{\gamma}}^{}_{i}X^{1} + \eta^{}_{R_{i}}, \nonumber \\ \forall i = 1,2,\cdots,N,
\label{eqn1} \\
y^{}_{D_{1}} \ = \ \sqrt{P^{0}_{s}} {\alpha^{}_{0}}X^{0} + \sqrt{P^{1}_{s}} {\alpha^{}_{0}}X^{1} + \eta^{}_{D_{1}}, \label{eqn2} \\
y^{}_{E_{1j}} \ = \ \sqrt{P^{0}_{s}} {\beta^{}_{0j}}X^{0} + \sqrt{P^{1}_{s}} {\beta^{}_{0j}}X^{1} + \eta^{}_{E_{1j}}, \nonumber \\ \forall j=1,2,\cdots,J_{}, 
\label{eqn3} \\
y^{}_{D_{2}} \ = \ \boldsymbol{\alpha}^{} {{\boldsymbol{\phi}^{}}}X^{0} + \boldsymbol{\alpha}^{} {{\boldsymbol{\psi}^{}}}X^{1} + \eta^{}_{D_{2}},
\label{eqn4} \\
y^{}_{E_{2j}} \ = \ \boldsymbol{\beta}^{}_j {{\boldsymbol{\phi}}^{}}X^{0} + \boldsymbol{\beta}^{}_j {{\boldsymbol{\psi}}^{}}X^{1} + \eta^{}_{E_{2j}}, \nonumber \\ \forall j=1,2,\cdots,J_{}. 
\label{eqn5}
\end{eqnarray}
The noise components, $\eta$'s, are assumed to be iid $\mathcal{CN}(0,N_0)$.
%\vspace{-2mm}
\section{Beamforming with Secret and Non-Secret Messages - Known CSI on all Links}
\label{sec3}
In this section, we assume perfect knowledge of the CSI on all links. This 
assumption can be valid in scenarios where the eavesdroppers are also 
legitimate users in the network. Since the symbol $X^{0}$ is transmitted at 
information rate $R_{0}$ irrespective of $X^{1}$, treating $X^{1}$ as noise, 
relays will be able to decode $X^{0}$ if $\forall i = 1,2,\cdots,N,$
\begin{eqnarray}
\frac{1}{2} I\big(X^{0}; \ y^{}_{R_{i}}\big)  =  \frac{1}{2} \log_{2}\Big(1 + \frac{P^{0}_{s}\mid\gamma_{i}\mid^{2}}{N_{0} + P^{1}_{s}\mid\gamma_{i}\mid^{2}}\Big) \ \geq \ R_{0}, \label{eqn6}
\end{eqnarray}
where (\ref{eqn6}) is derived using (\ref{eqn1}) and the factor $1/2$ appears 
because of the two hops. Similarly, using (\ref{eqn2}) and (\ref{eqn4}), the 
destination $D$ will be able to decode $X^{0}$ if 
\begin{eqnarray}
\frac{1}{2} I\big(X^{0}; \ y^{}_{D_{1}}, \ y^{}_{D_{2}}\big)  =  \frac{1}{2} \log_{2}\Big(1 + \frac{P^{0}_{s}\mid\alpha_{0}\mid^{2}}{N_{0} + P^{1}_{s}\mid\alpha_{0}\mid^{2}} + \nonumber \\ \frac{\boldsymbol{\phi}^{\ast}\boldsymbol{\alpha}^{\ast}\boldsymbol{\alpha}^{}\boldsymbol{\phi}^{}}{N_{0} + \boldsymbol{\psi}^{\ast}\boldsymbol{\alpha}^{\ast}\boldsymbol{\alpha}^{}\boldsymbol{\psi}^{}} \Big) \ \geq \ R_{0}. \label{eqn7}
\end{eqnarray}
Using (\ref{eqn1}) and with the knowledge of the symbol $X^{0}$, the information 
rate for $X^{1}$ at the $i$th relay is
\begin{eqnarray}
\frac{1}{2} I\big({X^{1}; \ y^{}_{R_{i}}}{\mid}{X^{0}}\big) \ = \ \frac{1}{2} \log_{2}\Big(1 + \frac{P^{1}_{s}\mid\gamma_{i}\mid^{2}}{N_{0}}\Big). \label{eqn8}
\end{eqnarray}
Similarly, using (\ref{eqn2}) and (\ref{eqn4}), the information rate for $X^{1}$ 
at the destination $D$ is
\begin{eqnarray}
\frac{1}{2} I\big({X^{1}; \ y^{}_{D_{1}}, \ y^{}_{D_{2}}}{\mid}{X^{0}}\big) \ = \ \frac{1}{2} \log_{2}\Big(1 + \frac{P^{1}_{s}\mid\alpha_{0}\mid^{2}}{N_{0}} + \nonumber \\ \frac{\boldsymbol{\psi}^{\ast}\boldsymbol{\alpha}^{\ast}\boldsymbol{\alpha}^{}\boldsymbol{\psi}^{}}{N_{0}} \Big). \label{eqn9} 
\end{eqnarray}
Using (\ref{eqn3}), (\ref{eqn5}), and assuming the knowledge of $X^0$ at the 
eavesdroppers, the information rate for $X^{1}$ at eavesdropper $E_{j}$ is
\begin{eqnarray}
\frac{1}{2} I\big({X^{1}; \ y^{}_{E_{1j}}, \ y^{}_{E_{2j}}}{\mid}{X^{0}}\big) = \frac{1}{2} \log_{2}\Big(1 + \frac{P^{1}_{s}\mid\beta_{0j}\mid^{2}}{N_{0}} + \nonumber \\ \frac{\boldsymbol{\psi}^{\ast}\boldsymbol{\beta}^{\ast}_{j}\boldsymbol{\beta}^{}_{j}\boldsymbol{\psi}^{}}{N_{0}} \Big). \label{eqn10} 
\end{eqnarray}
We note that there is no decoding constraint for the symbol $X^{0}$ on any 
eavesdropper $E_{j}$ similar to (\ref{eqn6}) and (\ref{eqn7}). This makes 
(\ref{eqn10}) as the best possible information rate for symbol $X^{1}$ at $E_{j}$.
Further, with the knowledge of symbol $X^{0}$, the relays will be able to 
decode the symbol $X^{1}$ if $\forall i = 1,2,\cdots,N$ \cite{ic8, ic12}
\begin{eqnarray}
\frac{1}{2} I\big({X^{1}; y^{}_{R_{i}}}{\mid}{X^{0}}\big) \geq
\frac{1}{2} I\big({X^{1}; y^{}_{D_{1}}, y^{}_{D_{2}}}{\mid}{X^{0}}\big), 
\end{eqnarray}
i.e.,
\begin{eqnarray}
\frac{1}{2} \log_{2}\Big(1 + \frac{P^{1}_{s}\mid\gamma_{i}\mid^{2}}{N_{0}}\Big) \geq \frac{1}{2} \log_{2}\Big(1 + \frac{P^{1}_{s}\mid\alpha_{0}\mid^{2}}{N_{0}} + \nonumber \\ \frac{\boldsymbol{\psi}^{\ast}\boldsymbol{\alpha}^{\ast}\boldsymbol{\alpha}^{}\boldsymbol{\psi}^{}}{N_{0}} \Big). \label{eqn11}
\end{eqnarray}
The constraint on the total transmit power is 
\begin{eqnarray}
P^{0}_{s} + P^{1}_{s} + \boldsymbol{\phi}^{\ast} \boldsymbol{\phi}^{} + \boldsymbol{\psi}^{\ast} \boldsymbol{\psi}^{} \quad \leq \quad P_T.  \label{eqn12}
\end{eqnarray}
Subject to the constraints in (\ref{eqn6}), (\ref{eqn7}), (\ref{eqn11}), and 
(\ref{eqn12}), the worst case achievable secrecy rate for $X^{1}$ is obtained 
by solving the following optimization problem \cite{ic50, ic8, ic12}:
\begin{eqnarray}
\hspace{-4mm}
R^{}_s = \max_{P^{0}_{s}, \ P^{1}_{s}, \atop{\boldsymbol{\phi}, \ \boldsymbol{\psi}}} \ \min_{j: 1,2,\cdots,J}\Big\{ \frac{1}{2} I\big({X^{1}; y^{}_{D_{1}}, y^{}_{D_{2}}}{\mid}{X^{0}}\big) -  \nonumber \\ \frac{1}{2} I\big({X^{1}; y^{}_{E_{1j}}, y^{}_{E_{2j}}}{\mid}{X^{0}}\big)\Big\}^{+}, \label{eqn13} 
\end{eqnarray}
s.t.
\begin{eqnarray}
\frac{1}{2} I\big(X^{0}; \ y^{}_{R_{i}}\big) \ \geq \ R_{0}, \ \forall i = 1,2,\cdots,N,  \label{eqn14}\\
\frac{1}{2} I\big(X^{0}; y^{}_{D_{1}}, y^{}_{D2}\big) \ \geq \ R_{0}, \label{eqn15} \\
\frac{1}{2} I\big({X^{1}; y^{}_{R_{i}}}{\mid}{X^{0}}\big) \ \geq \ \frac{1}{2} I\big({X^{1}; y^{}_{D_{1}}, y^{}_{D_{2}}}{\mid}{X^{0}}\big), \nonumber \\
\forall i = 1,2,\cdots,N, \label{eqn16} \\
P^{0}_{s} \ \geq \ 0, \ P^{1}_{s} \ \geq \ 0, \ P^{0}_{s} + P^{1}_{s} + \boldsymbol{\phi}^{\ast} \boldsymbol{\phi}^{} + \boldsymbol{\psi}^{\ast} \boldsymbol{\psi}^{} \ \leq \ P_T, \label{eqn17}
\end{eqnarray}
where ${\{a\}}^{+} = \max (a, 0)$, and without loss of generality 
we drop this operator since secrecy rate is non-negative.
The constraints (\ref{eqn14}), (\ref{eqn15}), and (\ref{eqn16}) are obtained from 
(\ref{eqn6}), (\ref{eqn7}), and (\ref{eqn11}), respectively. The objective function 
in (\ref{eqn13}) is obtained from (\ref{eqn9}) and (\ref{eqn10}). 
We solve the optimization problem in (\ref{eqn13}) as follows. 

$\bf{Step1:}$ Divide the total available transmit power $P_{T}$ in $M$ 
discrete steps of size $\Delta_{P_{T}} = \frac{P_{T}}{M}$, and let 
$P_m = m \Delta_{P_{T}}$, where $m = 0, 1, 2,\cdots,M-1$. 

$\bf{Step2 :}$ Rewrite the optimization problem (\ref{eqn13}) as 
the following two separate optimization problems; Problem 1 and 
Problem 2. 

\vspace{2mm}
{\bf Problem 1:} 
\begin{eqnarray}
R^{m}_s  \ = \nonumber \\ \max_{P^{1}_{s}, \ \boldsymbol{\psi}} \ \min_{j: 1,2,\cdots,J} \frac{1}{2} \Big\{ \log_{2}\Big(1 + \frac{P^{1}_{s}\mid\alpha_{0}\mid^{2} + \boldsymbol{\psi}^{\ast}\boldsymbol{\alpha}^{\ast}\boldsymbol{\alpha}^{}\boldsymbol{\psi}^{}}{N_{0}} \Big) \nonumber \\ 
- \log_{2}\Big(1 + \frac{P^{1}_{s}\mid\beta_{0j}\mid^{2}+ \boldsymbol{\psi}^{\ast}\boldsymbol{\beta}^{\ast}_{j}\boldsymbol{\beta}^{}_{j}\boldsymbol{\psi}^{}}{N_{0}} \Big)\Big\}^{}, 
%\nonumber \\ 
\label{eqn18} 
\end{eqnarray}
s.t.
\begin{eqnarray}
\ \forall i = 1,2,\cdots,N, \quad \frac{1}{2} \log_{2}\Big(1 + \frac{P^{1}_{s}\mid\gamma_{i}\mid^{2}}{N_{0}}\Big) \quad \geq \nonumber \\ \frac{1}{2} \log_{2}\Big(1 + \frac{P^{1}_{s}\mid\alpha_{0}\mid^{2} + \boldsymbol{\psi}^{\ast}\boldsymbol{\alpha}^{\ast}\boldsymbol{\alpha}^{}\boldsymbol{\psi}^{}}{N_{0}} \Big),  \nonumber \\
P^{1}_{s} \ \geq \ 0, \quad P^{1}_{s} + \boldsymbol{\psi}^{\ast} \boldsymbol{\psi}^{} \ \leq \ P_m. %\nonumber \\ 
\label{eqn19}
\end{eqnarray}
The optimization problem in (\ref{eqn18}) is a function of $P^{1}_{s}$, 
$\boldsymbol{\psi}$, and $P_{m}$. For a given $P_{m}$, it can be 
solved using semi-definite relaxation technique in \cite{ic12}. 

\vspace{2mm}
{\bf Problem 2:}
\begin{eqnarray}
\text{find} \ \ P^{0}_{s}, \ \boldsymbol{\phi}, \label{eqn20}
\end{eqnarray}
{\small
\begin{eqnarray}
\text{s.t.} \quad \forall i = 1,2,\cdots,N, \nonumber \\
\frac{1}{2} \log_{2}\Big(1 + \frac{P^{0}_{s}\mid\gamma_{i}\mid^{2}}{N_{0} + P^{1}_{s}\mid\gamma_{i}\mid^{2}}\Big) \geq R_{0}, \  \nonumber \\
\frac{1}{2} \log_{2}\Big(1 + \frac{P^{0}_{s}\mid\alpha_{0}\mid^{2}}{N_{0} + P^{1}_{s}\mid\alpha_{0}\mid^{2}} + \frac{\boldsymbol{\phi}^{\ast}_{}\boldsymbol{\alpha}^{\ast}\boldsymbol{\alpha}^{}\boldsymbol{\phi}^{}_{}}{N_{0} + \boldsymbol{\psi}^{\ast}\boldsymbol{\alpha}^{\ast}\boldsymbol{\alpha}^{}\boldsymbol{\psi}^{}} \Big) \geq R_{0}, \nonumber \\
P^{0}_{s} \geq 0, \ P^{0}_{s} + \boldsymbol{\phi}^{\ast} \boldsymbol{\phi} \leq P_T - P_{m}. \label{eqn21}
\end{eqnarray}
}

\vspace{1mm}
For a given $P^{1}_{s}$, $\boldsymbol{\psi}$, and $P_{m}$, 
it is obvious that the optimum direction of $\boldsymbol{\phi}$ which minimizes the
transmit power $P^{0}_s + \boldsymbol{\phi}^{\ast}\boldsymbol{\phi}$, subject to the constraints in (\ref{eqn21}),
lies in the direction of $\boldsymbol{\alpha}^{\ast}$, i.e., 
$\boldsymbol{\phi} = \sqrt{P^{0}_{R}}\boldsymbol{\phi}_{u}$, where 
$\boldsymbol{\phi}_{u} = \frac{\boldsymbol{\alpha}^{\ast}}{\parallel\boldsymbol{\alpha}^{\ast}\parallel}$ 
and $P^{0}_{R}$ is the relays transmit power associated with $X_{0}$. 
With this, we rewrite the feasibility problem in (\ref{eqn20}) in the following form:
% \vspace{-1mm}
% {\small
\begin{eqnarray}
\text{find} \ \ P^{0}_{s}, \ P^{0}_{R}, \label{eqn26}
\end{eqnarray}
% }
% \vspace{1mm}
% \hspace{-2mm}
s.t.
% \vspace{-1mm}
% {\small
\begin{eqnarray}
\Big(1 + \frac{P^{0}_{s}\mid\gamma_{i}\mid^{2}}{N_{0} + P^{1}_{s}\mid\gamma_{i}\mid^{2}}\Big) \ \geq \ 2^{2R_{0}}, \ \forall i = 1,2,\cdots,N, \nonumber \\
\Big(1 + \frac{P^{0}_{s}\mid\alpha_{0}\mid^{2}}{N_{0} + P^{1}_{s}\mid\alpha_{0}\mid^{2}} + \frac{P^{0}_{R}\boldsymbol{\phi}^{\ast}_{u}\boldsymbol{\alpha}^{\ast}\boldsymbol{\alpha}^{}\boldsymbol{\phi}^{}_{u}}{N_{0} + \boldsymbol{\psi}^{\ast}\boldsymbol{\alpha}^{\ast}\boldsymbol{\alpha}^{}\boldsymbol{\psi}^{}} \Big) \ \geq \ 2^{2R_{0}}, \nonumber \\
P^{0}_{s} \geq 0, \ P^{0}_{R} \geq 0, \ P^{0}_{s} + P^{0}_{R} \leq P_T - P_{m}. 
\label{eqn22}
\end{eqnarray}
% }
% \vspace{1mm}
% \hspace{-3mm}
For a given $P^{1}_{s}$, $\boldsymbol{\psi}$, and $P_{m}$, the feasibility problem in (\ref{eqn26})
with its constraints in (\ref{eqn22}) is a linear feasibility 
problem in $P^{0}_{s}$ and $P^{0}_{R}$, and it can be easily solved using 
linear programming techniques.

It can be shown that the secrecy rate $R^{m}_{s}$ which is obtained by solving 
the optimization problem (\ref{eqn18}) for a given $P_m$ is a strictly increasing 
function in $P_m$ \cite{ic12}. Hence, the idea is to find the maximum power $P_{m}$
for which $P^{1}_{s}$ and $\boldsymbol{\psi}$ obtained by solving (\ref{eqn18}) 
also gives a feasible solution $P^{0}_{s}$ and $P^{0}_{R}$ in (\ref{eqn26}) 
satisfying the constraints in (\ref{eqn22}). This can be achieved by decreasing 
$m$ from $M-1$ towards 0 and finding the maximum $m$ for which the solution of 
the optimization problem (\ref{eqn18}) (i.e., $P^{1}_{s}$, $\boldsymbol{\psi}$) 
with $P_{m}$ as available power also gives a feasible solution for (\ref{eqn26})
(i.e., $P^{0}_{s}$ and $P^{0}_{R}$ 
satisfying the constraints in (\ref{eqn22})).
\subsection{Suboptimal beamforming with non-secret message for $D$ and all $E_{j}$s}
In this subsection, we give a suboptimal beamforming method with secret and non-secret messages where
the secret message $W_{1}$ is intended only for $D$ whereas the non-secret message $W_{0}$ is intended 
for $D$ as well as all $E_{j}$s. The non-secret message is transmitted at a fixed rate $R_{0}$.
Similar to (\ref{eqn7}), using (\ref{eqn3}), (\ref{eqn5}) and treating $X^{1}$ as noise, $E_{j}$s
will be able to decode $X^{0}$ if $\forall j = 1,2,\cdots,J,$
\begin{eqnarray}
\frac{1}{2} I\big(X^{0}; \ y^{}_{E_{1j}}, \ y^{}_{E_{2j}}\big) = \frac{1}{2} \log_{2}\Big(1 + \frac{P^{0}_{s}\mid\beta_{0j}\mid^{2}}{N_{0} + P^{1}_{s}\mid\beta_{0j}\mid^{2}} \nonumber \\ +  \frac{\boldsymbol{\phi}^{\ast}\boldsymbol{\beta}^{\ast}_{j}\boldsymbol{\beta}^{}_{j}\boldsymbol{\phi}^{}}{N_{0} + \boldsymbol{\psi}^{\ast}\boldsymbol{\beta}^{\ast}_{j}\boldsymbol{\beta}^{}_{j}\boldsymbol{\psi}^{}} \Big) \ \geq \ R_{0}.\label{eqn27}
\end{eqnarray}
With this, the optimization problem in (\ref{eqn13}) will have additional constraints (\ref{eqn27}).
Similarly, the feasibility problem (\ref{eqn20}) will have the additional constraints (\ref{eqn27}). 
For a given $P^{1}_{s}$, $\boldsymbol{\psi}$, and $P_{m}$, 
the optimum direction of $\boldsymbol{\phi}$ which minimizes the
transmit power $P^{0}_s + \boldsymbol{\phi}^{\ast}\boldsymbol{\phi}$, can
be obtained by solving the following optimization problem:
\begin{eqnarray}
\min_{P^{0}_{s}, \ \boldsymbol{\Phi}} \ P^{0}_s + trace(\boldsymbol{\Phi}), \label{eqn28} \\
% \end{eqnarray}
% \begin{eqnarray}
\text{s.t.} \quad \quad \forall i = 1,2,\cdots,N, \nonumber \\
\Big(1 + \frac{P^{0}_{s}\mid\gamma_{i}\mid^{2}}{N_{0} + P^{1}_{s}\mid\gamma_{i}\mid^{2}}\Big) \ \geq \ 2^{2R_{0}},  \nonumber \\
\Big(1 + \frac{P^{0}_{s}\mid\alpha_{0}\mid^{2}}{N_{0} + P^{1}_{s}\mid\alpha_{0}\mid^{2}} + \frac{\boldsymbol{\alpha}^{}\boldsymbol{\Phi}^{}_{}\boldsymbol{\alpha}^{\ast}}{N_{0} + \boldsymbol{\psi}^{\ast}\boldsymbol{\alpha}^{\ast}\boldsymbol{\alpha}^{}\boldsymbol{\psi}^{}} \Big) \ \geq \ 2^{2R_{0}}, \nonumber \\
\Big(1 + \frac{P^{0}_{s}\mid\beta_{0j}\mid^{2}}{N_{0} + P^{1}_{s}\mid\beta_{0j}\mid^{2}} + \frac{\boldsymbol{\beta}^{}_{j}\boldsymbol{\Phi}^{}_{}\boldsymbol{\beta}^{\ast}_{j}}{N_{0} + \boldsymbol{\psi}^{\ast}\boldsymbol{\beta}^{\ast}_{j}\boldsymbol{\beta}^{}_{j}\boldsymbol{\psi}^{}} \Big) \ \geq \ 2^{2R_{0}}, \nonumber \\
\forall j = 1,2,\cdots,J, \ \boldsymbol{\Phi} \succeq \boldsymbol{0}, \ rank(\boldsymbol{\Phi}) = 1, \nonumber \\
P^{0}_{s} \geq 0, \ P^{0}_{s} + trace(\boldsymbol{\Phi}) \leq P_T - P_{m}, 
\label{eqn29}
\end{eqnarray}
where $\boldsymbol{\Phi} = \boldsymbol{\phi} \boldsymbol{\phi}^{\ast}$ and the constraints in (\ref{eqn29}) 
are written using all the constraints in (\ref{eqn21}) and (\ref{eqn27}). This is a non-convex optimization
problem which is difficult to solve. However, by relaxing the $rank(\boldsymbol{\Phi}) = 1$ constraint,
the above problem can be solved using semi-definite programming techniques. 
But, the solution $\boldsymbol{\Phi}$ of the above rank relaxed optimization problem may not have rank 1.
So, we take the largest eigen direction of $\boldsymbol{\Phi}$ as the suboptimal unit norm direction $\boldsymbol{\phi}_{u}$.
We substitute $\boldsymbol{\phi}_{u}$ in the feasibility problem (\ref{eqn26}) and its constraints
(\ref{eqn22}) and additional constraints (\ref{eqn27}). The remaining procedure to find 
$P^{0}_{s}, \ P^{1}_{s}, \ P^{0}_{R}$ and $\boldsymbol{\psi}_{}$ remains same as discussed in $\bf{Step 1}$ and
$\bf{Step 2}$.
\section{Beamforming with Secret and Non-Secret Messages -- Statistical CSI 
on Eavesdroppers Links}
\label{sec4}
In this section, we obtain the source and relays powers under the assumption 
that only the statistical knowledge of the eavesdroppers CSI is available. 
The eavesdropper CSI is assumed to be iid 
$\mathcal{CN}(0,\sigma^2_{\beta^{}_{0j}})$ for the direct link from 
source to $E_j$ and iid $\mathcal{CN}(0,\sigma^2_{\beta^{}_{ij}})$ for 
the link from relay $i$ to $E_j$. With this statistical knowledge of the 
eavesdroppers CSI, the optimization problem (\ref{eqn18}) can be 
written in the following form:
\begin{eqnarray}
%\hspace{-4mm}
R^{m}_s = \nonumber \\ \max_{P^{1}_{s}, \ \boldsymbol{\psi}} \ \min_{j: 1,2,\cdots,J} \frac{1}{2} \Big\{ \log_{2}\Big(1 + \frac{P^{1}_{s}\mid\alpha_{0}\mid^{2} + \boldsymbol{\psi}^{\ast}\boldsymbol{\alpha}^{\ast}\boldsymbol{\alpha}^{}\boldsymbol{\psi}^{}}{N_{0}} \Big) \nonumber \\ 
- \mathbb{E}\Big[\log_{2}\Big(1 + \frac{P^{1}_{s}\mid\beta_{0j}\mid^{2}+ \boldsymbol{\psi}^{\ast}\boldsymbol{\beta}^{\ast}_{j}\boldsymbol{\beta}^{}_{j}\boldsymbol{\psi}^{}}{N_{0}} \Big) \Big] \Big\}^{} \\
\label{eqn23} 
\geq \nonumber \\
\max_{P^{1}_{s}, \ \boldsymbol{\psi}} \ \min_{j: 1,2,\cdots,J} \frac{1}{2} \Big\{ \log_{2}\Big(1 + \frac{P^{1}_{s}\mid\alpha_{0}\mid^{2} + \boldsymbol{\psi}^{\ast}\boldsymbol{\alpha}^{\ast}\boldsymbol{\alpha}^{}\boldsymbol{\psi}^{}}{N_{0}} \Big) \nonumber \\ 
-  \log_{2}\Big(1 + \frac{P^{1}_{s}\sigma^2_{\beta^{}_{0j}}+ \boldsymbol{\psi}^{\ast}\Lambda_{\boldsymbol{\beta}^{}_{j}}\boldsymbol{\psi}^{}}{N_{0}} \Big)\Big\}^{}, 
\label{eqn24}
\end{eqnarray}
\begin{eqnarray}
\text{s.t.} \quad \forall i = 1,2,\cdots,N, \quad \frac{1}{2} \log_{2}\Big(1 + \frac{P^{1}_{s}\mid\gamma_{i}\mid^{2}}{N_{0}}\Big) \ \geq \nonumber \\ \frac{1}{2} \log_{2}\Big(1 + \frac{P^{1}_{s}\mid\alpha_{0}\mid^{2} + \boldsymbol{\psi}^{\ast}\boldsymbol{\alpha}^{\ast}\boldsymbol{\alpha}^{}\boldsymbol{\psi}^{}}{N_{0}} \Big),  \nonumber  \\
P^{1}_{s} \ \geq \ 0, \quad P^{1}_{s} + \boldsymbol{\psi}^{\ast} \boldsymbol{\psi}^{} \quad \leq \quad P_m, \label{eqn25}
\end{eqnarray}
where the lower bound 
in (\ref{eqn24}) is due to Jensen's inequality. The 
$\Lambda_{\boldsymbol{\beta}^{}_{j}}$ in (\ref{eqn24}) is a diagonal matrix with 
$[\sigma^2_{\beta^{}_{1j}},\sigma^2_{\beta^{}_{2j}},\cdots,\sigma^2_{\beta^{}_{Nj}}]^{T}$
on its diagonal. For a given $P_{m}$, the optimization problem (\ref{eqn24}) can 
be solved using semi-definite relaxation. The optimal 
$P^{0}_{s}$, $P^{1}_{s}$, $P^{0}_{R}$, and $\boldsymbol{\psi}$ can be obtained
by solving the optimization problems (\ref{eqn24}) and (\ref{eqn26}) as discussed 
in Section-\ref{sec3}.
\section{Results and Discussions}
\label{sec5}
We present the numerical results and discussions in this section. We 
obtained the secrecy rate results through simulations for $N = 2$ relays 
and $J = 1,2,3$ eavesdroppers. The following complex channel gains are 
taken in the simulations:
$\alpha_{0} = 0.3039 + 0.5128i$, 
$\beta_{01} = 0.1161 - 0.0915i$, 
$\beta_{02} = -0.5194 + 0.4268i$,  
$\beta_{03} = -0.0900 + 0.4769i$,  
$\boldsymbol{\gamma} = [-1.3136 + 0.3534i, \ -0.7070 - 1.1305i]^{}$, 
$\boldsymbol{\alpha} = [0.3241 + 0.4561i, \  0.2713 - 0.5850i]^{}$,  
$\boldsymbol{\beta}_{1} = [-0.6407 + 0.0709i, \ -0.0562 + 0.5120i]^{}$, 
$\boldsymbol{\beta}_{2} = [0.1422 - 0.6060i, \ -0.0590 - 0.3308i]^{}$,
and 
$\boldsymbol{\beta}_{3} = [0.2793 - 0.1426i, \  -0.5092 + 0.2570i]^{}$.
For the case of statistical CSI on eavesdroppers links, the following
parameters are taken:
$\sigma^2_{\beta^{}_{01}} = 0.01$, 
$\sigma^2_{\beta^{}_{02}} = 0.04$, 
$\sigma^2_{\beta^{}_{03}} = 0.09$, 
$\sigma^2_{\beta^{}_{i1}} = 0.25$, 
$\sigma^2_{\beta^{}_{i2}} = 0.36$, 
$\sigma^2_{\beta^{}_{i3}} = 0.49$, $i=1,2$. 
The value of $M$ is taken to be 50.

\begin{figure}[htb]
\begin{minipage}[b]{.48\linewidth}
\centering
\centerline{\epsfig{figure=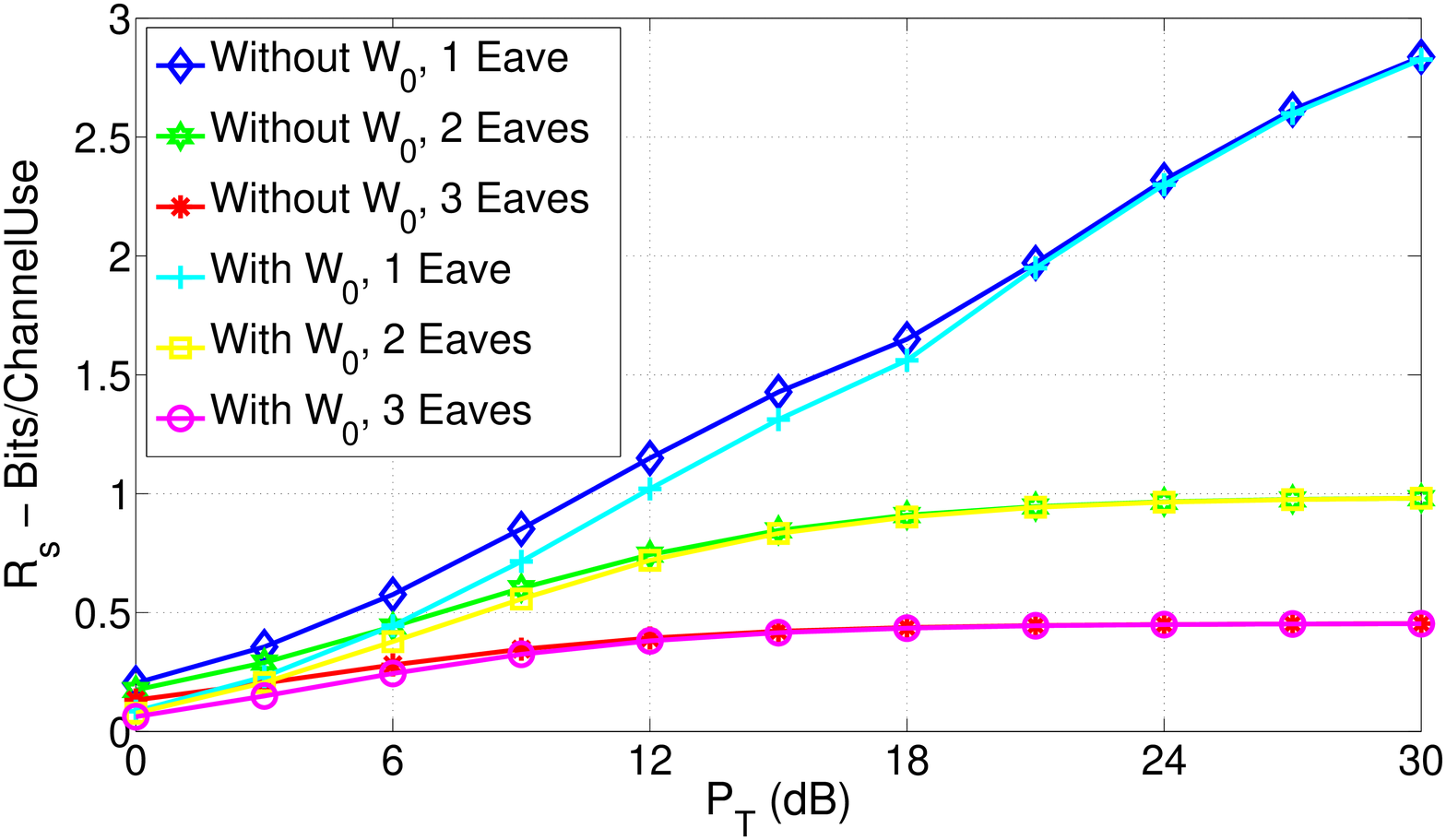,width=4.85cm,height=7.5cm}}
\centerline{\small{(a)}}\medskip
\end{minipage}
\hfill
\begin{minipage}[b]{0.48\linewidth}
\centering
\centerline{\epsfig{figure=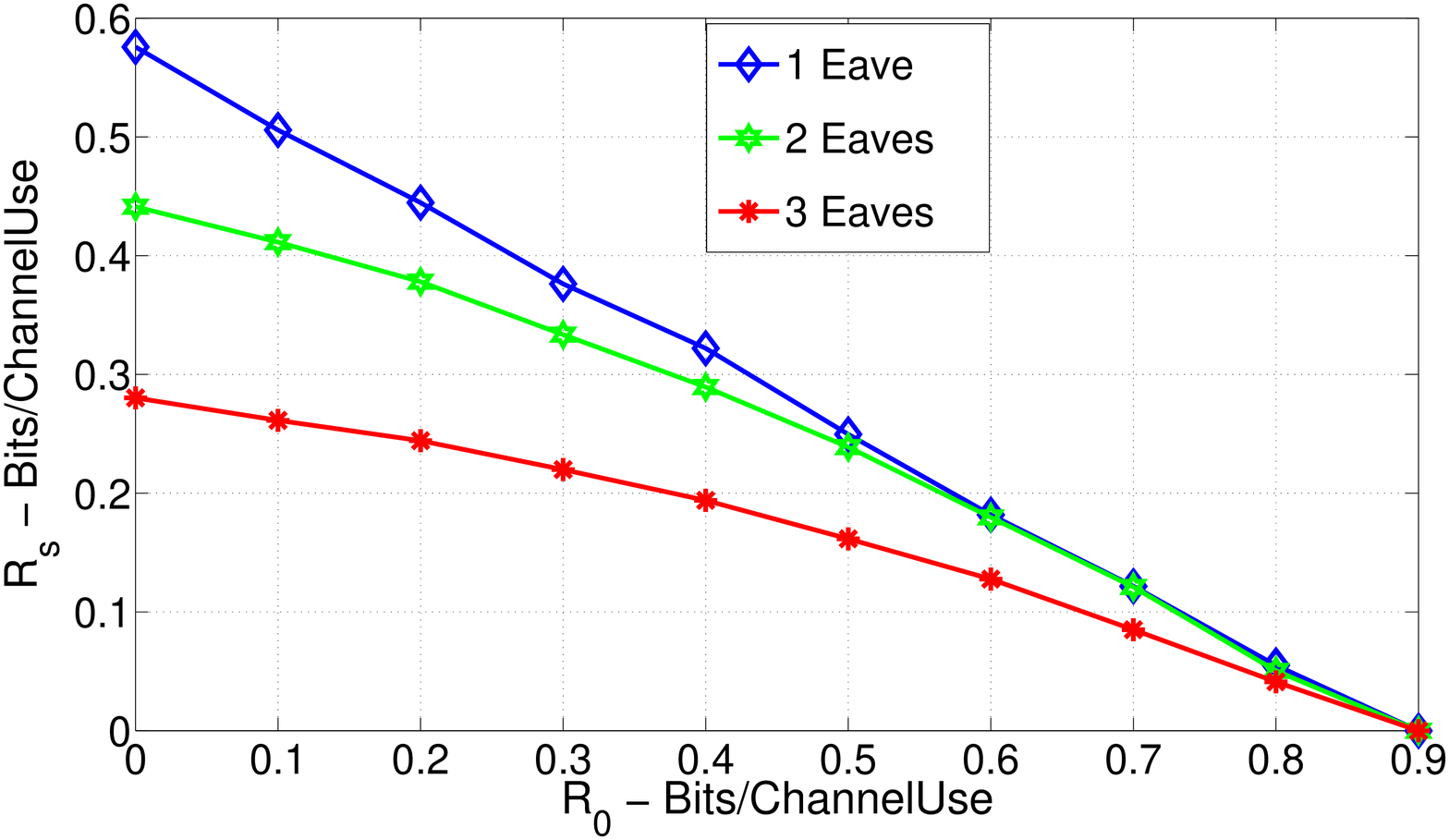,width=4.85cm, height=7.5cm}}
\centerline{\small{(b)}}\medskip
\end{minipage}
\caption{Secrecy rate of DF relay beamforming for $N=2$, $J=1,2,3$.
Perfect CSI on all links. (a) secrecy rate vs total power ($P_T$) with 
and without $W_{0}$, $R_0=0.2$; (b) {$R_s$ vs $R_0$} for 
$P_T=6$ dB.}
\label{fig2}
\end{figure}

{\em Perfect CSI on all Links:}
Figure \ref{fig2}(a) shows the secrecy rate plots for DF relay beamforming 
as a function of total transmit power ($P_T$) for the case when perfect CSI
on all links is assumed. The secrecy rates are plotted for the cases of 
with and without $W_{0}$ for 2 relays and different number of eavesdroppers. 
For the case with $W_{0}$, the information rate of the $W_{0}$ is fixed at 
$R_0=0.2$. We also assume that when $W_{0}$ is present, it is intended only for $D$
and it need not be protected from $E_{j}$s. 
From Fig. \ref{fig2}(a), we observe that, for a given number of 
eavesdroppers, the secrecy rate degrades when $W_{0}$ is present. However, 
this degradation becomes insignificant when $P_T$ is increased to large 
values. Also, the secrecy rate degrades for 
increasing number of eavesdroppers. Figure \ref{fig2}(b) shows the $R_s$ 
vs $R_0$ tradeoff, where $R_s$ is plotted as a function of $R_0$ for 
$J=1,2,3$ at a fixed total power of $P_T=6$ dB. It can be seen that as 
$R_0$ is increased, secrecy rate decreases. This is because the available 
transmit power for $W_{1}$ decreases as $R_0$ is increased. 
In Fig. \ref{fig2}(b), we see that the maximum achievable secrecy 
rate $R_{s}$ without $W_{0}$ (i.e., when $R_0=0$), which we denote by $R_s'$, 
are $0.58$, $0.45$ and $0.28$ for $J = 1,2,3$ eavesdroppers, respectively. It 
can be further noted that if $R_{0} \leq R_{s}'$, then $W_{0}$ can also be 
transmitted as a secret message and the remaining rate $R_{s}'-R_{0}$ can be 
used for the secret message ($W_{1}$) transmission. In other words, if 
$R_{0} \leq R_{s}'$, then it is possible for both $W_1$ and $W_0$ to be sent 
as secret massage at a combined secrecy rate $R_s'$. However, if $R_{0}>R_{s}'$, 
then $W_{0}$ can not be transmitted as a secret message.

\begin{figure}[htb]
\begin{minipage}[b]{.48\linewidth}
\centering
\centerline{\epsfig{figure=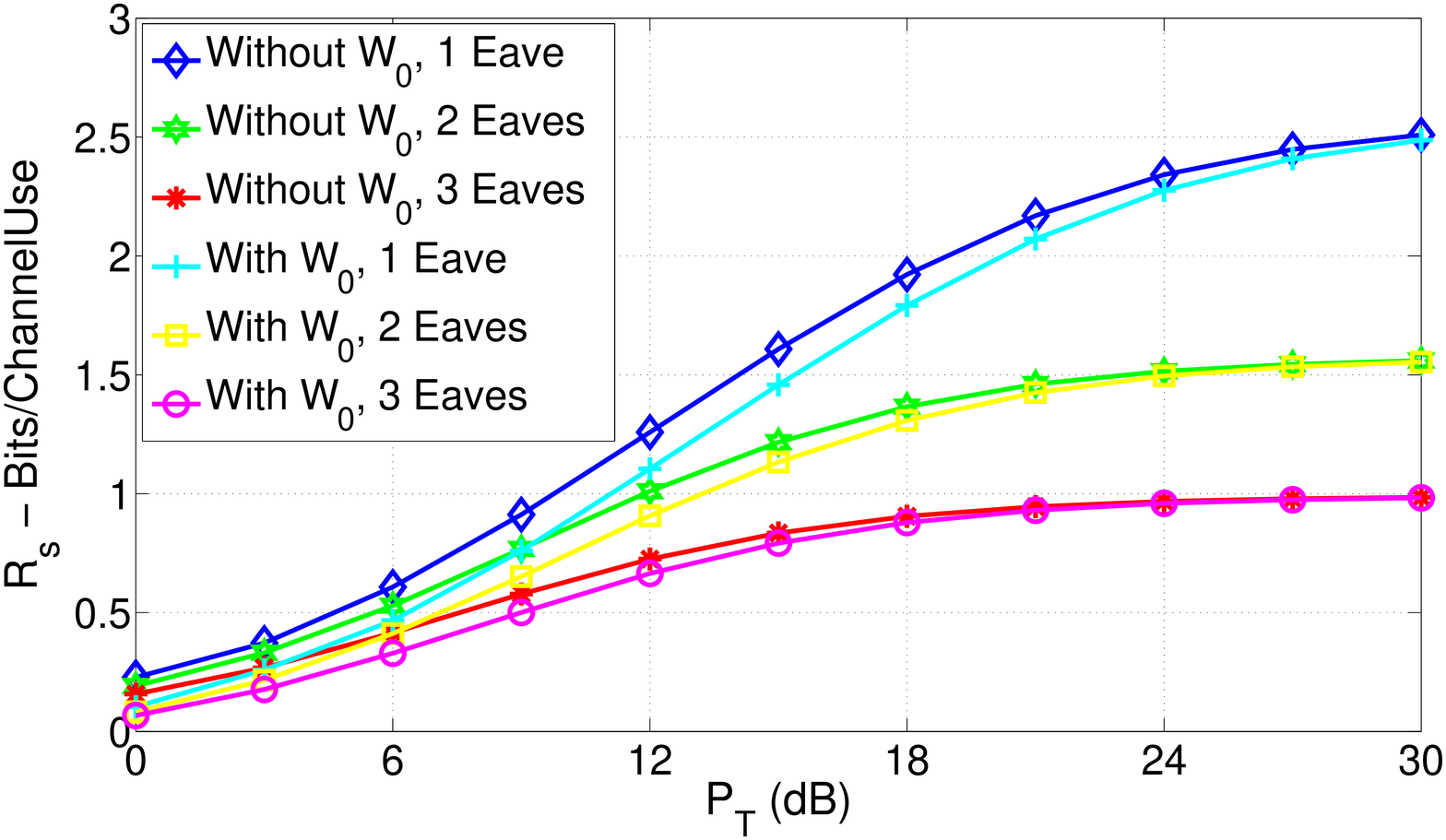,width=4.85cm,height=7.5cm}}
\centerline{\small{(a)}}\medskip
\end{minipage}
\hfill
\begin{minipage}[b]{0.48\linewidth}
\centering
\centerline{\epsfig{figure=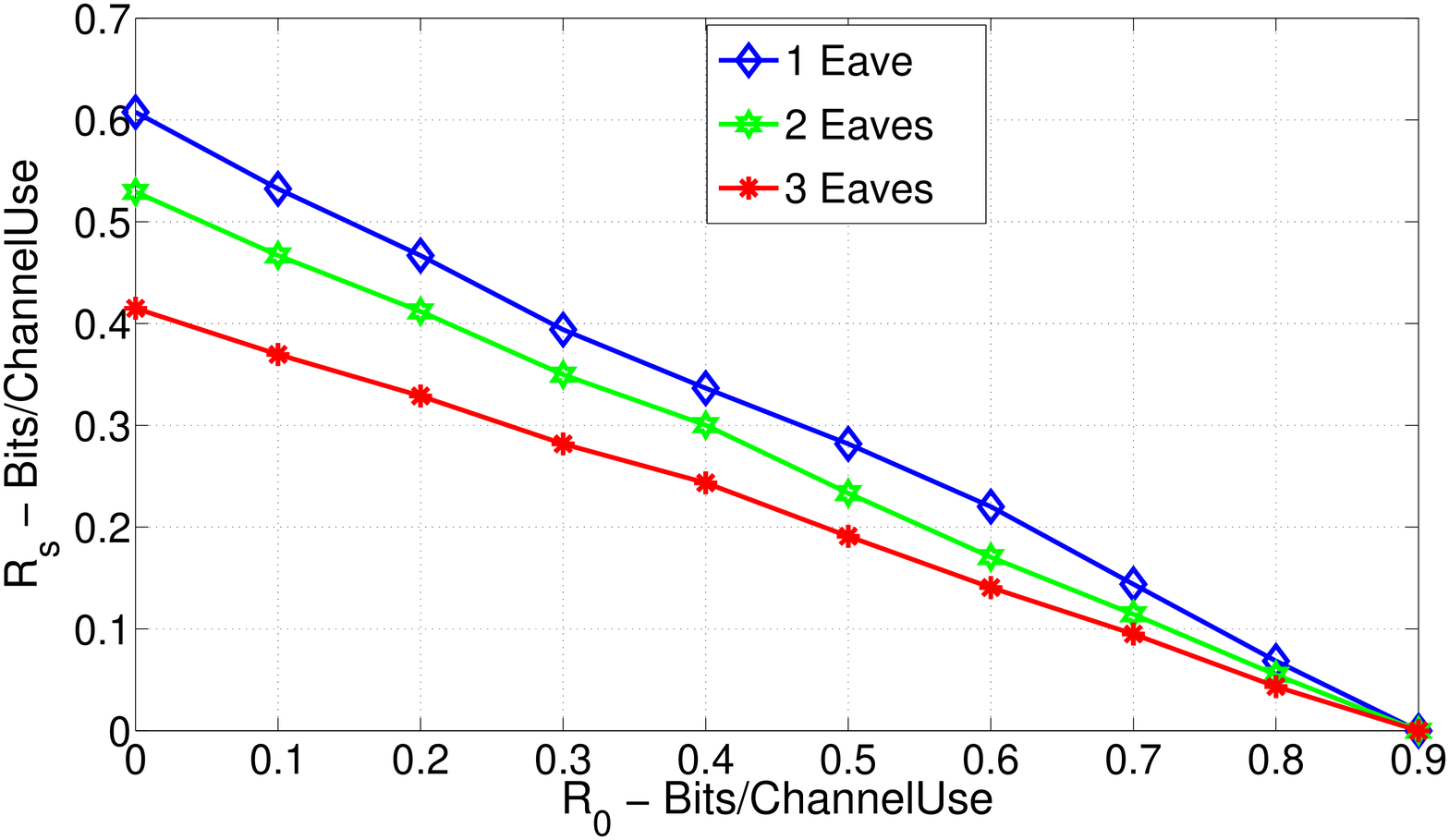,width=4.85cm, height=7.5cm}}
\centerline{\small{(b)}}\medskip
\end{minipage}
\caption{Secrecy rate of DF relay beamforming for $N=2$, $J=1,2,3$.
Statistical CSI on eavesdroppers links. (a) secrecy rate vs $P_T$ with 
and without $W_{0}$, $R_0=0.2$; (b) {$R_s$ vs $R_0$} for 
$P_T=6$ dB.}
\label{fig4}
\end{figure}

{\em Statistical CSI on eavesdroppers links:}
Figures \ref{fig4}(a) and (b) show the secrecy rate plots for DF relay beamforming 
for the case when only the statistical CSI on eavesdroppers links is assumed to be
known. The CSI on the other links are assumed to be perfectly known. Figure
\ref{fig4}(a) shows the secrecy rate versus $P_T$ plots for $R_0=0.2$, and
Fig. \ref{fig4}(b) shows the secrecy rate versus $R_0$ plots for $P_T=6$ dB.
Observations similar to those in the case of perfect CSI on all links are
observed in Figs. \ref{fig4}(a) and \ref{fig4}(b) as well. 
\section{Conclusions}
\label{sec6}
We investigated beamforming in DF relaying 
using multiple relays, where the source sends a secret message as well 
as a non-secret message to the destination node in the presence of multiple 
non-colluding eavesdroppers. 
The source and relays operate under a total power constraint. 
We obtained the optimum source powers and weights 
of the relays for both secret and non-secret messages which maximized the worst 
case secrecy rate for the secret message as well as met the information rate 
constraint $R_{0}$ for the non-secret message. 
We solved this problem for the 
cases when ($i$) perfect CSI of all links was known, 
and ($ii$) only the statistical CSI of the eavesdroppers links and perfect CSI 
of other links were known.

\end{document}